\documentclass[5p,twocolumn,times,number]{elsarticle}

\usepackage{graphicx}
\usepackage{amsmath}   

\begin{document}

\begin{frontmatter}

\title{Commissioning of the ATLAS Liquid Argon Calorimeter}

\author[add1]{S.~Laplace\corref{cor}}
\ead{laplace@lapp.in2p3.fr}
\author{\\on behalf of the ATLAS Liquid Argon Calorimeter Group}

\cortext[cor]{Corresponding author}

\address[add1]{LAPP, Universit\'e de Savoie, CNRS/IN2P3, Annecy-Le-Vieux, France}

\begin{abstract}
The in-situ commissioning of the ATLAS liquid argon calorimeter is taking place since
three years. During this period, it has been fully tested by means of
frequent calibration runs, and the analysis of the large cosmic muon
data samples and of the few beam splash events that occurred on September
10th, 2008. This has allowed to obtain a stable set of calibration
constants for the first collisions, and to measure the in-situ
calorimeter performances that were found to be at the expected level. 
\end{abstract}


\end{frontmatter}

\section{Introduction}

The construction and installation of the liquid argon (LAr) calorimeter and
its readout system were completed in March 2006. Since then, the
calibration and readout systems have been extensively used by taking
very frequent calibration runs. Physics data coming from cosmic muons
(since summer 2006 up to now) and from the first LHC beam events
(September 10 to 12th, 2008) were analyzed and used to measure
the calorimeter in-situ performance.

\section{The Liquid Argon Calorimeter}

The ATLAS LAr calorimeters consist of four calorimeters located in
three cryostats filled with liquid argon which acts as active
medium~\cite{:2008zzm}. The passive material and the geometry are
specific to each part, and are detailed below:  

\begin{itemize}
\item the electromagnetic barrel and endcap calorimeters (EMB and
  EMEC) provide a precise measurement of electron and photon positions
  and energies up to a pseudo rapidity of 3.2. Their absorbers are
  made of lead, achieving a minimal radiation length of 22 $X_0$. Their
  specific accordion geometry ensures a full $\phi$ hermiticity and a uniform and
  fast response. They are segmented in three longitudinal compartments
  (called the strip, middle and back samplings) to
  extract the shower shape, as well as a presampler layer in 
  order to estimate the loss due to the dead material in front of the
  calorimeter. The resolution is expected to be $\sigma(E)/E = 10 \%
  /\sqrt{E} \oplus 0.7\%$. 
\item the hadronic endcap (HEC) is a classical sandwich calorimeter
  with copper as passive material. Its pseudo rapidity coverage ranges
  from 1.5 to 3.2 and it is also segmented in four longitudinal
  compartments. The resolution for hadrons is expected to be $\sigma(E)/E = 50 \%
  /\sqrt{E} \oplus 3\%$. 
\item the forward calorimeter (FCAL) detects the particles in the
  forward region with a pseudo rapidity coverage between 3.2 and
  4.8. Due to the high particles occupancy in this region, a specific
  geometry with very thin liquid argon gaps (between 250m and
  500m) has been adopted to limit the space charge, what could induce
  detection inefficiencies. The absorbers are made of tungsten (in the
  first compartment) or copper (in the second and third
  compartments). The resolution for hadrons is expected to be $\sigma(E)/E = 100 \%
  /\sqrt{E} \oplus 10\%$. 
\end{itemize}


The three sub-calorimeters share the same readout electronics: the
signal read from the electrodes is first gathered into cells (a cell
gathers four electrodes along the $\phi$ direction) by means of
summing boards located on mother boards. It is then processed by the Front End
Board~\cite{Buchanan:2008zzc} (FEB) hosted just outside the cryostats. The raw
triangular ionization signal is amplified and split into three linear
gain scales in the ratio 1/10/100. To optimize the signal-to-noise
ratio, the signal is shaped by a bipolar $CR-(RC)^2$ filter. It is
then sampled at the LHC bunch-crossing frequency of $40$ MHz and
stored in pipelines during the L1 latency. For
events accepted by the L1 trigger, five samples in one gain scale are
read from the pipeline and digitized by 12-bit ADCs, to be finally
sent by optical fibers to the back-end
electronics~\cite{Bazan:2007zz} housed 70 m away from the detector. 

The back-end system digitally processes the data coming from the
FEB. It performs various data integrity checks and higher level
monitoring tasks, but most importantly, it applies an optimal
filtering algorithm to the samples in order to compute the energy and
timing of every calorimeter cell.


A calibration signal~\cite{Colas:2008zz} with an exponential shape can be injected to the
mother boards in order to monitor the electronics
response and to compute the electronic gain. Because the injected
calibration signal has a different shape than the ionization one
(exponential versus triangular), and because it is injected at a
different point (mother boards versus electrodes), the pulse shape
that is measured at the FEB exit differs in calibration and physics
modes. Several methods~\cite{collard,prieur,marco} allow to modelize
this difference by predicting the ionization pulse shape starting from
the calibration one: it is this predicted pulse shape that is used to
compute optimal filtering coefficients. 

\section{Calorimeter Performances}

\subsection{High energy deposits}

\begin{figure}[hbtp]
\centering
\includegraphics[width=0.95\linewidth]{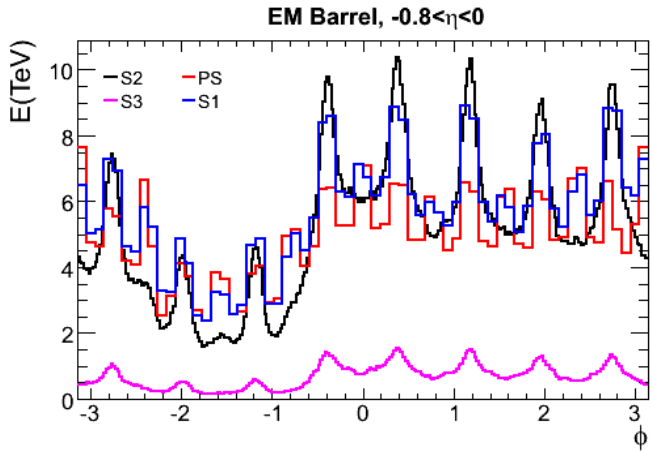}
\caption{Energy deposition in the LAr calorimeter barrel in each
  sampling (PS = presampler, S1 = front, S2 = middle, S3 = back)
  projected along the $\phi$ axis. } 
\label{fig:HighEnergy}
\end{figure}

The first beam splash events have been used to check high energy
depositions in the LAr calorimeter. Figure~\ref{fig:HighEnergy} shows
the projected energy along the $\phi$ axis for $-0.8<\eta<0$ for the
different LAr barrel calorimeter layers. Only cells with an energy
above $5\sigma$ above noise are considered. The eight-fold structure
reflects the endcap toroid matter at high radius (S2 and S3), and the
sixteen-fold structure shows the additional matter and shielding at
low radius (PS and S1). 

\subsection{Timing Alignment}

\begin{figure}[hbtp]
\centering
\includegraphics[width=0.95\linewidth]{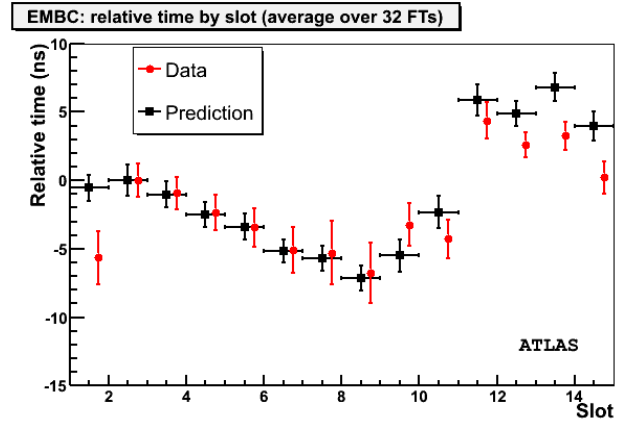}
\caption{Comparison between the predicted (black squares) and measured
(red dots) cell timings for each FEB slot in the front end crate, averaged
over all front-end crates. }
\label{fig:Timing}
\end{figure}

Beam splash events also allow to check the cell timing and derive
timing delays for the first collisions. Figure~\ref{fig:Timing} shows
the comparison between the predicted and measured cell timings averaged over
all front-end crates as a function of the FEB slot in the crate. The
measured timings are obtained using the optimal filtering coefficients
and are corrected by a time-of-flight correction to make as if the
particles were coming from the collision point. The predicted timings
are derived from the calibration timings taking into account the
different cable lengths involved in the readout path. The agreement
between the two timings is better than $\pm 2$ ns for most of the slots. The
residual disagreements can be corrected using a programmable delay on
each FEB.  

\subsection{Ionization Pulse Shape}

\begin{figure}[hbtp]
\centering
\includegraphics[width=0.95\linewidth]{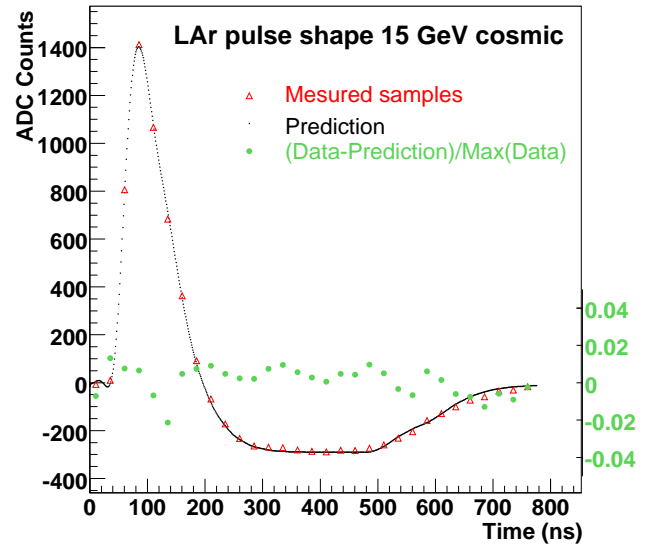}
\caption{Comparison between the predicted (black dots) and measured
  (red triangles) pulse shape samples in cosmic muons data. The green
  dots show the normalized difference between the two.}
\label{fig:PulseShape}
\end{figure}

The prediction of the ionization pulse shape has been successfully
validated in past testbeams. Cosmic muons and first LHC beam can also
be used to check this prediction in-situ. The comparison between the
predicted and measured pulse shapes in the electromagnetic LAr
calorimeter using cosmic muons is shown in
Fig.~\ref{fig:PulseShape}. An agreement at the level of $2\%$ is
observed in this randomly chosen cell, but a more global study shows
that the agreement ranges from $2\%$ in the barrel to $4\%$ in the
endcap calorimeter. 

\subsection{Calorimeter Uniformity}

\begin{figure}[hbtp]
\centering
\includegraphics[width=0.95\linewidth]{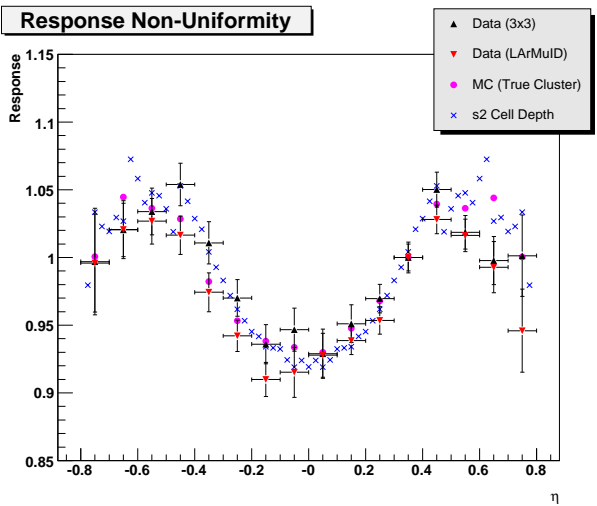}
\caption{Most probable value from the landau fit at different $\eta$
  positions: the black and red triangles show the data, reconstructed
  with two different clusters (see text). The red dots show the Monte
  Carlo response, while the blue crosses represent the average cell
  depth at a given $\eta$ position.}
\label{fig:Uniformity}
\end{figure}

The response uniformity~\cite{cooke} of the electromagnetic barrel calorimeter was
measured in-situ using cosmics muons data collected between August 2006
and March 2007. In these data, only nine calorimeter modules were
readout, and no inner detector information was available. The very
small energy deposition of the muon (only due to ionization) is reconstructed using two
different clusters: a so-called ``$3\times 3$'' cluster
gathering the cells in all samplings falling inside a square of
$3\times 3$ in middle cell unit around a seed cell, and a ``LArMuID'' cluster that
gathers a few cells based on signal-to-noise values. The energy
distributions of these two clusters are then fit with a Landau
function that accounts for the ionization energy deposition fluctuations,
convoluted with a Gaussian function that accounts for the detector
resolution. The Most Probable Value (MPV) of the Landau is plotted as
a function of $\eta$ for both Monte Carlo (MC) and data in
Fig.~\ref{fig:Uniformity}. The $10\%$ variation of the energy
deposition along $\eta$ is due to the different cell depths: 
since the muon energy deposition is proportionnal to the path length
in the calorimeter, the landau MPV naturally follows the cell depth
variations. The difference between the MC and data MPV distributions
quantifies the response uniformity which is found to be of the order
of $2\%$. 

\subsection{Missing Transverse Energy}

Random triggers can be used to check the missing transverse energy ($E_T^{miss}$)
resolution on events that should have no $E_T^{miss}$. The missing transverse
energy is computed using two noise-suppression methods: in the first
method, all calorimeter cells that have a signal that is $2\sigma$ above
noise are kept. Since this threshold is low, one expects to 
create a fake $E_T^{miss}$ by picking up the upward noise fluctuations. In the
second method, a more refined "4-2-0" topological cluster algorithm is used:
the clustering starts from cells which a signal-to-noise ratio greater
than $4$. All cells around the seed that have a signal-to-noise ratio
above $2$ are added to the cluster. Finally, all neighboring cells of
the cluster are also added (they correspond to the last $0$
signal-to-noise threshold of the "4-2-0" configuration). As shown on
Fig.~\ref{fig:$E_T^{miss}$}, which displays the $E_T^{miss}$ for these two methods, this
second algorithm leads to much less fake $E_T^{miss}$. 

\begin{figure}[hbtp]
\centering
\includegraphics[width=0.99\linewidth]{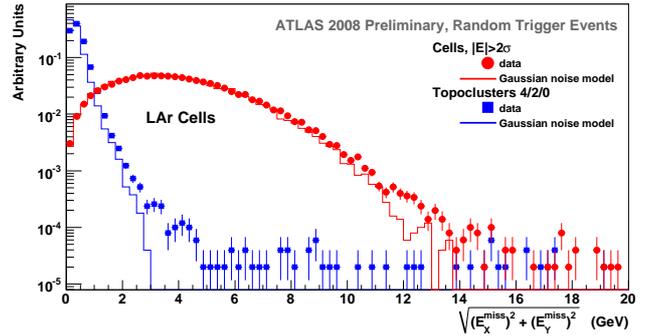}
\caption{Distributions of missing transverse energy for the two
  noise-suppression methods (see text). The measured points (dots) are
compared with the predicted $E_T^{miss}$ (full lines). }
\label{fig:$E_T^{miss}$}
\end{figure}

The measured $E_T^{miss}$ is compared to the expected distribution obtained from a
randomization of the cell energy with a Gaussian noise (for each cell,
energy values are picked by this Gaussian distribution which is
centered at 0 and has an standard deviation which is equal to the
respective $\sigma$ noise value). The measured $E_T^{miss}$ agrees well with
the expectation, except for the presence of a tail at larger $E_T^{miss}$
value, which has been understood and corrected since
then~\footnote{An unexpected source of noise coming from the filtering box
  on the high voltage lines of some PS modules was fixed.}.

\subsection{Electrons from Ionization}

A search for electrons coming from delta rays caused by cosmics muons
has been performed. To start with, $3.5$ million events that have a
track reconstructed at the trigger level 2 are considered. A loose $\phi$ 
track matching is performed with electromagnetic clusters that have a
transverse energy above $3$ GeV: $11000$  candidates remain at this
stage. The cluster lateral shower shapes in the first and
second calorimeter samplings are required to be consistent with those
from an electron. Finally, the associated track is required to 
have at least $25$ Transition Radiation Tracker (TRT)
hits~\footnote{The TRT is a drift-tube detector that is part of the
  ATLAS tracking system. It also provides particle identification
  between electrons and pions/muons by making use of transition
  radiation produced by relativistic particles such as electrons.},
and the remaining events are split in two categories:  
\begin{itemize}
\item 1229 events with only one reconstructed track which are expected
  to correspond to Bremsstrahlung photons coming from the muon.
\item 85 events with at least two reconstructed tracks which are considered as
  ionization electron candidates. 
\end{itemize}

Two variables are used to understand the background contamination
coming from the Bremsstrahlung events to the signal electron
candidates: the ratio $E/p$ of the energy $E$  reconstructed in the
calorimeter and the momentum $p$ reconstructed in the tracker ($E/p$
is around $1$ for electrons), and the ratio of high-threshold over the
low-threshold TRT hits, $N_{TRT}^{high/low}$, which is expected to be
larger for relativistic particles such as electrons than for other
non-relativistic particles such as muons. A signal box is defined in
these two parameters space, with limits that vary as a function of
$\eta$ and $p_T$. For example, at low $\eta$,  the limits are
$0.8<E/p<2.5$ and $N_{TRT}^{high/low}>0.8$. Less than $2\%$ (19 of the
1229) of the Bremsstrahlung candidates fall into the signal box, while
more than $40\%$ (36 out of 85) of the electron candidates satisfy the
same criteria. 

\begin{figure}[hbtp]
\centering
\includegraphics[width=0.95\linewidth]{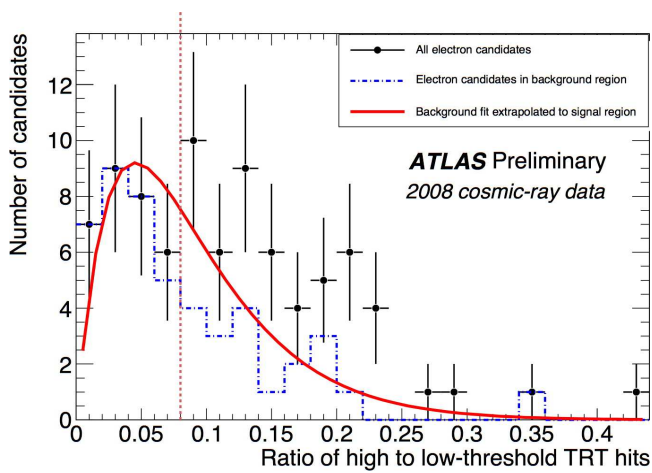}
\caption{Ratio of high-threshold over low-threshold TRT hits for all
  electron candidates (dots). The red line shows the expected
  background distribution.  }
\label{fig:Electrons}
\end{figure}

\begin{figure}[hbtp]
\centering
\includegraphics[width=0.95\linewidth]{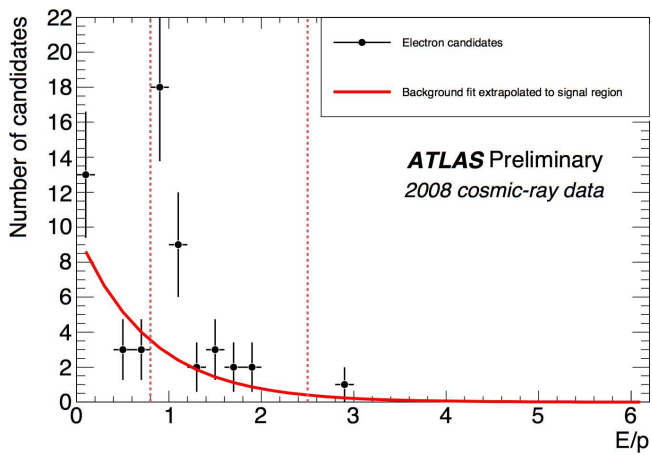}
\caption{$E/p$ for all electron candidates (dots). The red line
  shows the expected background distribution. }  
\label{fig:Electrons2}
\end{figure}

A two-dimensional maximum likelihood fit is performed over these two
variables to evaluate the background contribution to the signal box:
the projected distributions onto the two variables are shown on
Fig.~\ref{fig:Electrons} and~\ref{fig:Electrons2}. The background
probability distribution function is taken from the Bremsstrahlung
sample. Clear excesses are seen in the signal regions indicating the
presence of ionization electrons. From the fit, the number of
background events in the signal box is estimated to be $8.7 \pm 3.1$.  

\begin{figure}[hbtp]
\centering
\includegraphics[width=0.99\linewidth]{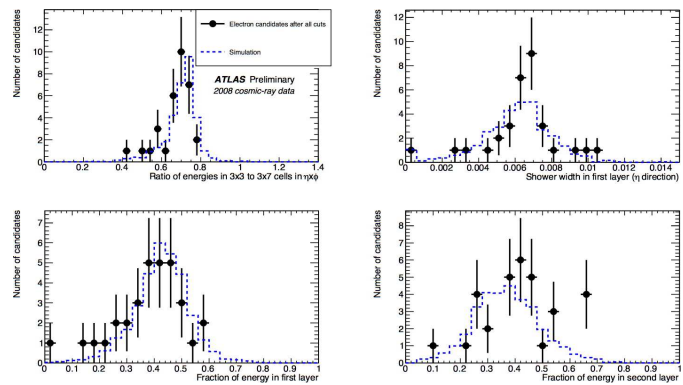}
\caption{Comparison of shower shapes between electron candidates in
  the cosmics data sample and a Monte Carlo simulation of projective
  electrons with a transverse energy of $5$ GeV. }
\label{fig:Electrons3}
\end{figure}

The events remaining in the signal box~\footnote{Over the 36 events
  remaining, 4 have a positive charge and are removed.} are further
studied: a comparison of the measured shower shapes and simulated
ones~\footnote{For that purpose, a Monte Carlo simulation of $5$ GeV
  projective electrons is used.} is shown on
Fig.~\ref{fig:Electrons3}: a good agreement between data and  Monte
Carlo is observed.  

\section{Conclusion}

The liquid argon calorimeters of ATLAS are installed in the cavern
since 2006. After 10 years of development, several beam tests and in
situ commissioning with cosmic muons, they were fully operational for
the first runs of LHC commissioning in September 2008. Results
obtained from cosmics muons and single beam events are promising, and
the first electrons were observed in the electromagnetic calorimeter. 

\section*{Acknowledgments}
The work presented here has been performed within the ATLAS LAr
collaboration. It would not have been possible without the dedicated
effort of many people in our LAr detector group and the ATLAS
collaboration over many years. I am especially indebted to people who
built, integrated and installed the LAr detectors in the ATLAS cavern
and those who operate the detector on a daily basis.


\end{document}